\documentclass[twocolumn,preprintnumbers,amsmath,amssymb,superscriptaddress]{revtex4}
\usepackage{bm}
\usepackage{ifpdf}
\usepackage{graphicx}
\usepackage{amssymb}
\usepackage{amsmath}
\usepackage{color}
\usepackage{hyperref}

\newcommand{\ket}[1]{\left|{#1}\right\rangle}
\newcommand{\bra}[1]{\left\langle{#1}\right|}

\newcommand{\braopket}[2]{\left\langle{#1}|{#2}|{#1}\right\rangle}

\newcommand{\beq}{\begin{equation}}
\newcommand{\eeq}{\end{equation}}

\newcommand{\jy}[1]{{#1}}

\newcommand{\bbsout}[1]{}
\newcommand{\blue}[1]{#1}
\newcommand{\bsout}[1]{}
\usepackage{siunitx}

\begin{document}

\title{Surpassing the no-cloning limit with a heralded hybrid linear amplifier  for coherent states}

\author{Jing~Yan~Haw}
\affiliation{Centre for Quantum Computation and Communication Technology, Department of Quantum Science, Research School of Physics and Engineering, The Australian National University, Canberra, ACT 2601, Australia}
\author{Jie~Zhao}
\affiliation{Centre for Quantum Computation and Communication Technology, Department of Quantum Science, Research School of Physics and Engineering, The Australian National University, Canberra, ACT 2601, Australia}
\author{Josephine Dias}
\affiliation{Centre for Quantum Computation and Communication
  Technology, School of Mathematics and Physics, University of
  Queensland, St. Lucia, Queensland 4072, Australia}
\author{Syed~M.~Assad}
\affiliation{Centre for Quantum Computation and Communication Technology, Department of Quantum Science, Research School of Physics and Engineering, The Australian National University, Canberra, ACT 2601, Australia}
 \author{Mark~Bradshaw}
\affiliation{Centre for Quantum Computation and Communication Technology, Department of Quantum Science, Research School of Physics and Engineering, The Australian National University, Canberra, ACT 2601, Australia}
\author{R\'{e}mi~Blandino}
\affiliation{Centre for Quantum Computation and Communication
  Technology, School of Mathematics and Physics, University of
  Queensland, St. Lucia, Queensland 4072, Australia}
\author{Thomas~Symul}
\affiliation{Centre for Quantum Computation and Communication Technology, Department of Quantum Science, Research School of Physics and Engineering, The Australian National University, Canberra, ACT 2601, Australia}
\author{Timothy~C.~Ralph}
\affiliation{Centre for Quantum Computation and Communication
  Technology, School of Mathematics and Physics, University of
  Queensland, St. Lucia, Queensland 4072, Australia}
\author{Ping~Koy~Lam}
\email{ping.lam@anu.edu.au}
\affiliation{Centre for Quantum Computation and Communication Technology, Department of Quantum Science, Research School of Physics and Engineering, The Australian National University, Canberra, ACT 2601, Australia}  

\begin{abstract}
The no-cloning theorem states that an unknown quantum state cannot be cloned exactly and deterministically due to the linearity of quantum
mechanics. Associated with
this theorem is the quantitative no-cloning limit that sets an
upper bound to the quality of the generated clones. However,
this limit can be circumvented by abandoning determinism and using probabilistic methods. 
Here, we report an experimental demonstration of probabilistic
cloning of arbitrary coherent states that clearly surpasses the no-cloning
limit. Our scheme is based on a hybrid linear amplifier that combines an
ideal deterministic linear amplifier with a
heralded measurement-based noiseless
amplifier. We demonstrate the production
of up to five clones with the fidelity of each clone clearly exceeding the corresponding
no-cloning limit. Moreover, since successful cloning events are heralded, our scheme has
the potential to be adopted in quantum repeater, teleportation and
computing applications.
\end{abstract}

\maketitle
The impossibility to perfectly duplicate an unknown quantum state deterministically, known as the no-cloning theorem~\cite{Wootters1982single}, lies at the heart of quantum information theory and guarantees the security of quantum cryptography~\cite{VScarani,Cerf2006optical}. This no-go theorem, however, does not rule out the possibility of imperfect cloning. The idea of generating approximate copies of an arbitrary quantum state was conceived by Buzek and Hillery in their seminal work~\cite{buvzek1996quantum} with the proposal of universal quantum cloning machine. This discovery has since sparked intense research in both
discrete~\cite{Gisin1997,Bruss1998,Buzek1998,Simon2000} and continuous variable~\cite{Cerf2000a,Cerf2000b,Lindblad2000,fiuravsek2001optical,Braunstein2001} systems to explore the fundamental limit of cloning fidelity allowed by quantum mechanics, known as the no-cloning limit. Several quantum cloning experiments approaching the optimal fidelity enforced by this limit have since been demonstrated for single photons~\cite{Lamas-Linares2002}, polarisation states~\cite{Fasel2002} and coherent states~\citep{Andersen2005}. 
\begin{figure*}[t]
\begin{center} 
\includegraphics[width = 13cm]{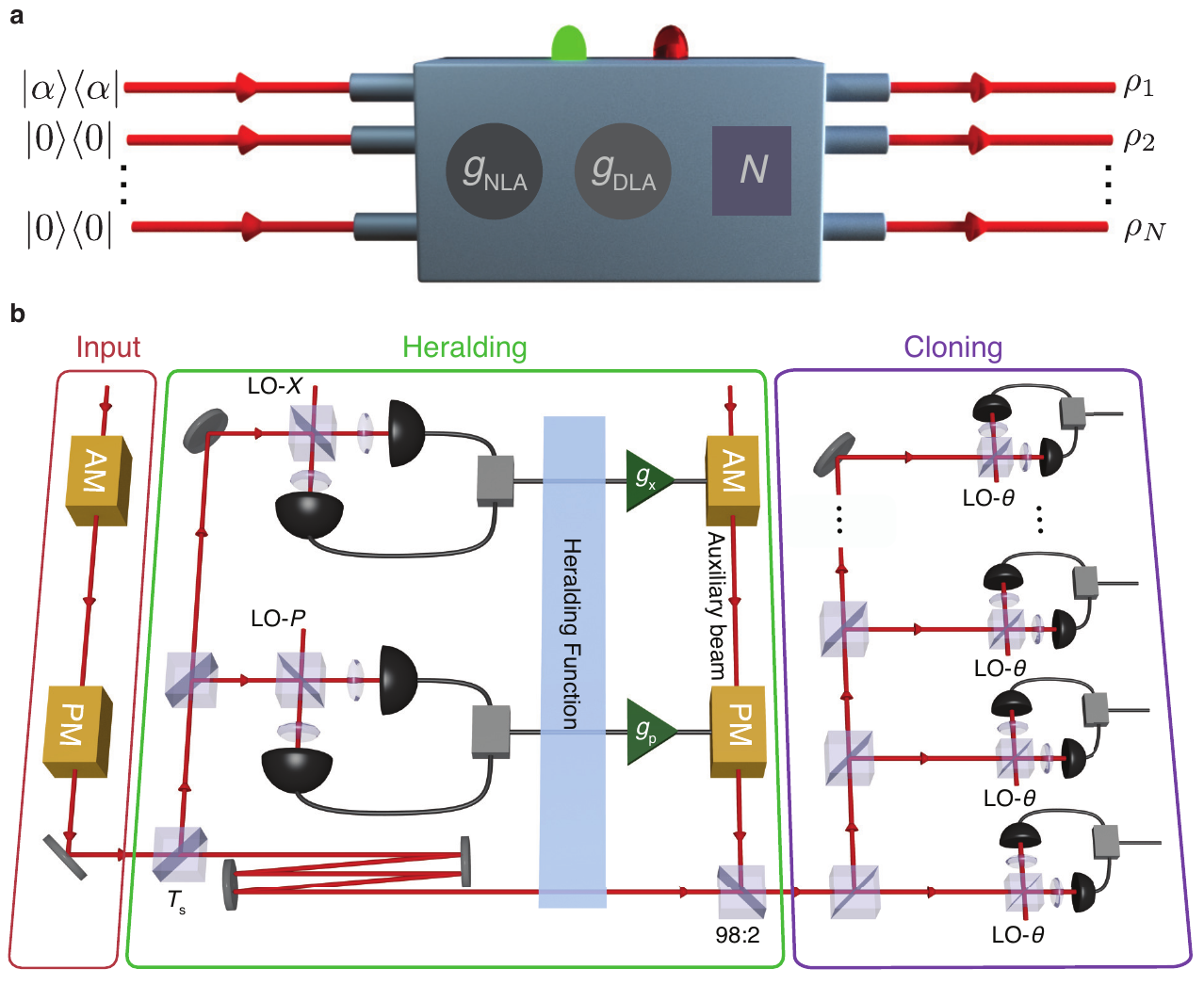}
\caption{\textbf{Hybrid Cloning Machine.} ({\bf a}) An $N$-port hybrid cloning machine (HCM), consisting
  of two control knobs: a probabilistic noiseless linear amplifier (NLA) gain ($g_\mathrm{NLA}$)
  and a deterministic linear amplifier (DLA) gain ($g_\mathrm{DLA}$). Heralded successful events (symbolised by a green light)
  produce $N$ clones ($\rho_{i}$) of coherent state $\ket{\alpha}$ 
  with noise less than the deterministic approach, while
  unsuccessful events (red light) will be discarded. ({\bf b}) Experimental
  schematic for HCM. When $g_\mathrm{NLA}<g_\mathrm{DLA}$, the cloning
  machine can be realised by a feed-forward scheme. The input
  coherent state passes through a beam splitter with transmitivity $T_\text{s}$, where both conjugate
  quadratures of the reflected port are measured via a dual-homodyne
  detection setup. The measurement outcomes pass through a heralding
  function and the successful events are then amplified with gain $g_\textrm{x,p}$ to displace the corresponding transmitted input state via a strong
  auxiliary beam. An $N$-port beam splitter finally creates $N$
  clones, which are characterised by homodyne measurements on quadratures $\theta=\{X,P\}$. $\ket{0}$, vacuum state; LO, local
  oscillator; 98:2, 98\% transmissive, 2\% reflective beam splitter; AM, amplitude modulator; PM, phase modulator. \label{fig:1}}
\end{center}
\end{figure*}
By forgoing determinism, perfect cloning is not entirely forbidden by the law of quantum physics. In fact, if the quantum states to be cloned are chosen from a discrete, linearly independent set, then the
unitarity of quantum evolution does allow probabilistic exact cloning~\cite{Duan1998,Ralph2003,VDunjko2012,CMCaves2013,Chen2011}. Non-deterministic high-fidelity cloning of linearly dependent input states can also be performed if the cloning operation is only arbitrarily close to the ideal case~\cite{Fiuravsek2004optimal,CMCaves2013}. Recently, the invention of probabilistic noiseless linear amplifier (NLA)~\cite{Ralph2009}, and its subsequent theoretical studies~\cite{marek2010coherent,fiuravsek2009engineering,jeffers2010nondeterministic,kim2012quantum,eleftheriadou2013quantum} and
implementations~\cite{Xiang2010,Ferreyrol2010,Zavatta2010,Usuga2010,donaldson2015experimental}
in principle provided a method for cloning arbitrary distributions of
coherent states with high fidelity via an amplify-and-split
approach~\cite{Xiang2010}. In practice, however, implementing NLA for coherent states with amplitude $|\alpha|\geq 1$ remains a technical challenge. This is because the resources required scales exponentially with the coherent state size.

In this article, we follow a different path by adopting a method that interpolates
between exact-probabilistic and approximate-deterministic cloning~\cite{chefles1999strategies}. We
show that a hybrid linear amplifier, comprising of a probabilistic NLA
and an optimal deterministic linear amplifier (DLA)~\cite{Andersen2005,andersen2009quantum}, followed by an $N$-port beam splitter is an effective quantum cloner. Previously, M{\"u}ller \emph{et~al.}~\cite{Muller2012} demonstrated probabilistic cloning of coherent states which outperformed the best deterministic scheme for input alphabet with random phases but fixed mean photon number. Here, we propose a high fidelity heralded cloning for
arbitrary distributions of coherent states and experimentally demonstrate the production of $N$ clones
with fidelity that surpasses the Gaussian no-cloning limit $F_{N}=N/(2N-1)$~\cite{fiuravsek2001optical,Braunstein2001}.
\section{Results}
\noindent \textbf{Hybrid Cloning Machine}. Our heralded hybrid cloning machine (HCM) is depicted conceptually in
Fig.~\ref{fig:1}a, where an $N$-copy cloner is parametrised
by an NLA amplitude gain ($g_\mathrm{NLA}$) and an optimal DLA gain
($g_\mathrm{DLA}$). By introducing an arbitrary input
coherent state of $|\alpha\rangle$, and setting the total gain to unity,
\begin{equation}
\label{eq:ug}
g=g_\mathrm{NLA}g_\mathrm{DLA}/\sqrt{N}=1,
\end{equation} 
HCM will generate $N$ clones with identical mean $\alpha$ and quadrature variance 
$1+2(g_\textrm{DLA}^2-1)/N$ (where the quantum noise level is 1). Since the
probabilistic amplification incurs no noise at all, the variance is a function of $g_\textrm{DLA}$ only. Such setup can be
interpreted as two linear amplifiers with distinct features,
complementing each other by sharing the burden of
amplification. Lower noise can be achieved at the expense of the probability of success by increasing the NLA gain. Conversely, a higher probability of success, though with an increased noise, can be obtained by increasing the DLA gain. Hence, by tailoring both gains appropriately, one can achieve the desired cloning fidelity, with vanishing probability of success as
the fidelity approaches unity.

A key feature in our implementation is the observation that when the
probabilistic gain is less than the deterministic gain, $g_\mathrm{NLA}<g_\mathrm{DLA}$, the hybrid
amplifier can be translated to a linear optical setup~\cite{andersen2009quantum} with an embedded
measurement-based NLA (MBNLA) (Fig.~\ref{fig:1}b). This equivalence is illustrated in Fig.~\ref{fig:sla} and discussed in more detail in Supplementary Note 1. The MBNLA is the post-selective version
of the physical realisation of NLA that has been
proposed~\cite{Fiurasek2012,Walk2013} and experimentally demonstrated 
recently~\cite{Chrzanowski2014}. Compared to its physical counterpart, MBNLA offers the ease of implementation and avoids the predicament of demanding experimental resources. By deploying MBNLA as the heralding function in a feed-forward control setup~\cite{RBlandino2016}, HCM preserves the amplified quantum state, extending the use of the MBNLA
beyond point-to-point protocols such as quantum key distribution.

\begin{figure}[t]
\centering
\includegraphics[width=6.8cm]{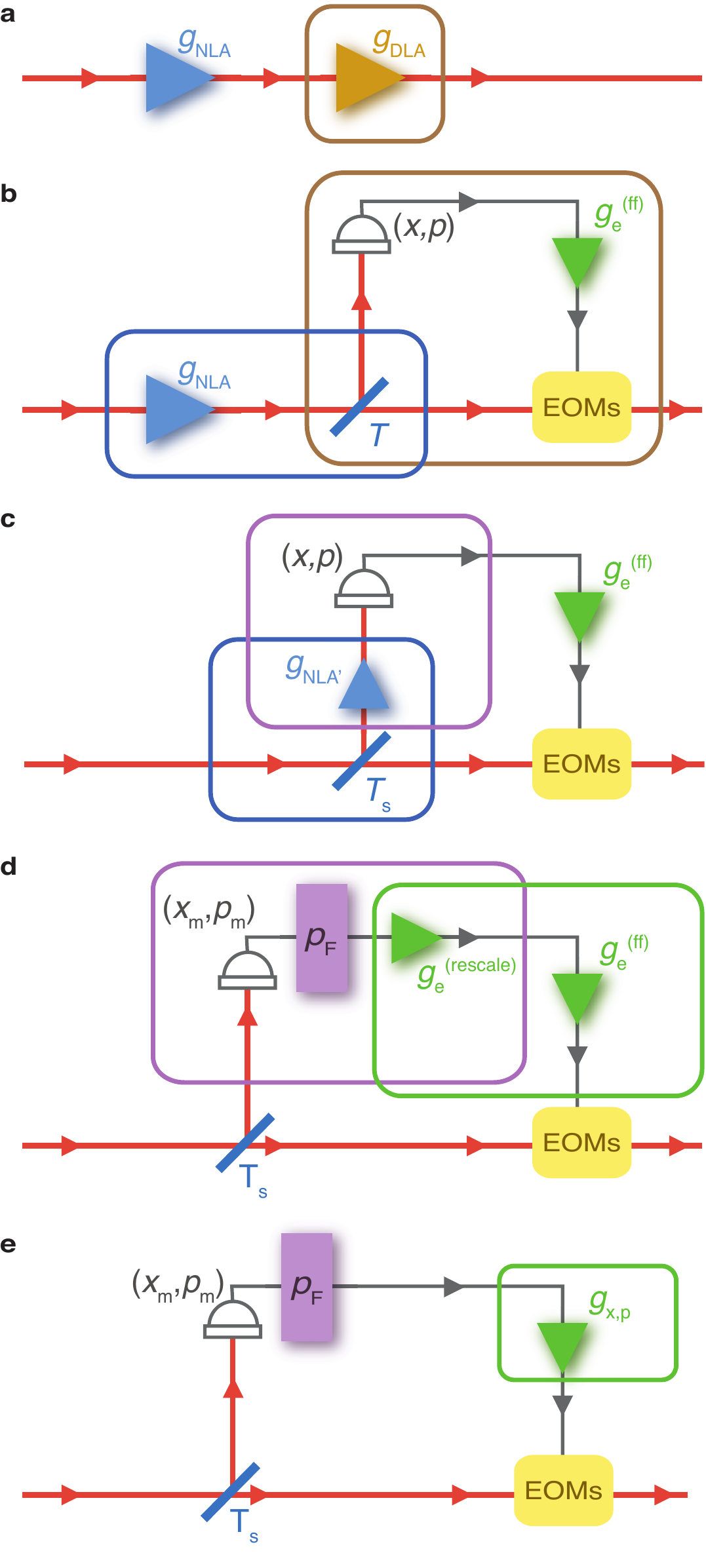}
\caption{\label{fig:sla} \textbf{Hybrid Linear Amplifier.} ({\bf a}) The general concatenated amplifier, consisting a noiseless linear amplifier (NLA) followed by a deterministic linear amplifier (DLA). ({\bf b}) An optical implementation of
  the DLA with a beam splitter of transmission $T$ and electronic gain $g_{\mathrm{e}}^{\mathrm{(ff)}}$. ({\bf c})
  When $g_{\mathrm{NLA}} < g_{\mathrm{DLA}}$, the NLA and beam splitter $T$ can
  be substituted by an effective NLA (NLA$'$) at the reflection port of a beam
  splitter $T_\text{s}$. ({\bf d}) The NLA$'$ followed by a dual-homodyne detection with outcomes $(x,p)$
  is replaced by a heralding function $p_\text{F}$ with an electronic rescaling
  $g_{\mathrm{e}}^{\mathrm{(rescale)}}$ acting upon measurement outcomes $(x_\text{m},p_\text{m})$. ({\bf e}) The two electronic gains are combined into $g_{\mathrm{x,p}}$. EOMs, electro-optical modulators.}
\end{figure} 

To clone an input coherent state, we first tap off part of the light with a beam splitter of transmission
\begin{equation}
\label{eq:ts}
T_\text{s}=(g_\mathrm{NLA}/g_\mathrm{DLA})^2,
\end{equation}
which is then detected on a
dual-homodyne detector setup locked to measure the amplitude and phase
quadratures ($X$ and $P$). 
\jy{A probabilistic heralding function, which is the probabilistic quantum filter function of an MBNLA~\cite{Fiurasek2012,Chrzanowski2014}, is then applied to the measurement outcome. By post selecting the dual-homodyne data with higher amplitude, the heralding function gives rise to an output distribution with higher overall mean. Mathematically, this function is given by
\begin{align}
\label{eq:filter}
  p_\text{F}\left(\alpha_\text{m} \right) =
\begin{cases}
  \frac{1}{M}\exp\left[|\alpha_\text{m}|^2\left(1-\frac{1}{g_{\text{NLA}'}^2} \right) \right]  &\mbox{if } |\alpha_\text{m}| < |\alpha_\text{c}|\\
  1 &\mbox{otherwise.}
\end{cases}
\end{align}
Here, $\alpha_\text{m}=(x_\text{m}+ip_\text{m})/\sqrt{2}$ is the dual-homodyne outcomes. $M=\exp \left[|\alpha_\text{c}|^2 \left(1-1/g_{\text{NLA}'}^2 \right) \right]$ is the normalization term with $\alpha_\text{c}$ as the NLA cut-off. The NLA gain $g_\text{NLA}'$ is tailored to achieve unity gain, while the cut-off $|\alpha_\text{c}|$ is chosen with respect to the gain and maximum input amplitude to emulate noiseless amplification faithfully while retaining a sufficient amount of data points (See Supplementary Note 2).}

The heralded signal is then scaled with gain
$g_\textrm{x,p}=\sqrt{2\left(1/T_\text{s}-1 \right)}$ and used to modulate an
auxiliary beam. The auxiliary beam is combined with the transmitted
beam using a 98:2 highly transmissive beam splitter, which acts as a displacement operator to the 
transmitted beam~\cite{paris1996displacement}. Finally, the combined beam passes through an
$N$-port beam splitter to create $N$-clones, which is then verified by
homodyne measurements.

\begin{figure*}[t]
\centering
\includegraphics[width=16cm]{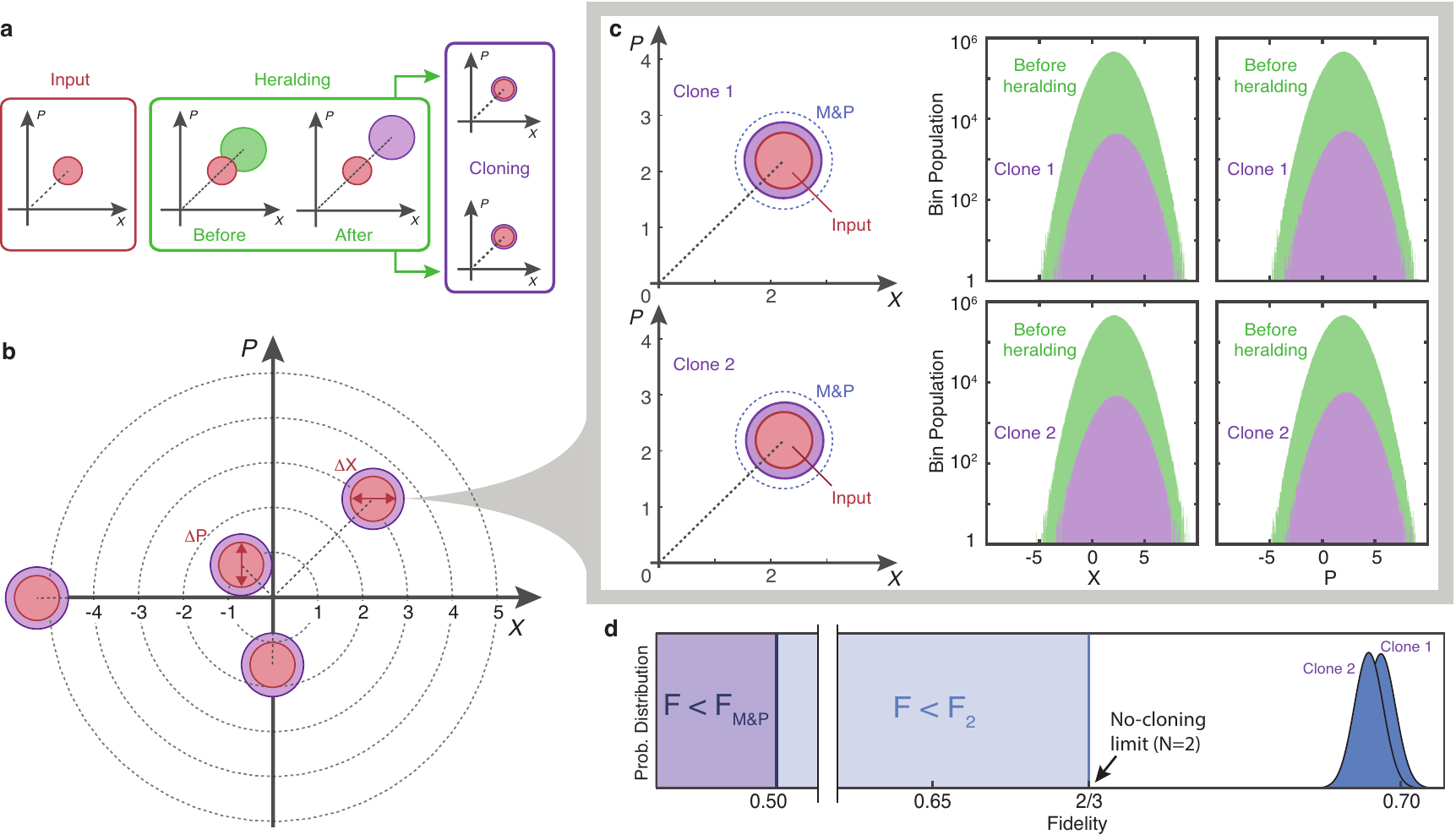}
\caption{\textbf{Two clones with hybrid cloning machine.} (\textbf{a}) Phase space
  representation of the deterministic-probabilistic hybrid cloning
  approach. The input state is first deterministically amplified
  before being heralded to produce a target state
  which is split into two clones. (\textbf{b}) Cloning of distinct
  input states. Since both the deterministic and noiseless linear amplifiers are invariant to the input
  state, any unknown coherent state can be cloned in the same
  way. For a coherent state, the quadrature standard deviations $\Delta X=\Delta P=1$. (\textbf{c}) Cloning of coherent state $(x,p)=\left(2.23, 2.19\right)$. Left, noise contours (1 standard deviation width) of the Wigner
  functions of the input state (red circle) and the clones from
  a measure-and-prepare (M\&P) cloning machine (dashed blue) and an hybrid cloning machine (purple circle). Right,
  Quadrature measurement histograms constructed from \num{5e7} homodyne measurements before (green) and  \num{5.9e5} measurements after heralding (purple). (\textbf{d}) Probability distributions of the fidelity of the clones. Both clones surpass the fidelity limits imposed by the
  M\&P cloner ($F_\mathrm{M\&P}=0.5$) and the deterministic cloner
  ($F_2=2/3$).\label{fig:2}}
\end{figure*}

In our experiment, the dual-homodyne measurement heralds successful operation shot-by-shot. This is then paired up with the
corresponding verifying homodyne measurements to select the
successful amplification events. The accumulated accepted
data points give the distribution of the conjugate quadratures of the successful clones. The processing of the input coherent
state at different stages of the HCM is illustrated by Fig.~\ref{fig:2}a.

It is instructive to compare our scheme to that of ref.~\cite{Muller2012}, where probabilistic cloning of fixed-amplitude coherent states was demonstrated. In~\cite{Muller2012}, the amplification is performed by a phase-randomized displacement and phase insensitive photon counting measurement, which have to be optimised according to input amplitude. In our scheme, the amplitude and the phase of the input state is
a priori unknown. Moreover, owing to the phase-sensitive dual-homodyne measurement and coherent feed-forward control in DLA~\cite{kim2012quantum,jeffers2011optical}, the state to be cloned is amplified coherently in the
desired quadrature. The integration of MBNLA in
HCM, which emulates a phase-preserving noiseless amplification, further enhances the amplitude of the signal while maintaining its phase.\\

\noindent \textbf{Two Clones.} To benchmark the performance of the HCM, Fig.~\ref{fig:2}b demonstrates the
universality of the cloning machine by showing the cloning results of four coherent input
states with different complex amplitudes $\ket{\alpha=x/2+ip/2}$, where
$(x,p)=(-0.71,0.72)$, $(-0.01,-1.51)$, $(2.23,2.19)$ and $(-5.26,-0.02)$. The
figure of merit we use is the fidelity
$F=\braopket{\alpha}{\rho_i}$, which quantifies the
overlap between the input state $\ket{\alpha}$ and the $i$-th clone $\rho_i$. Using a setting of $T_\text{s}\approx 0.6$, our device clones the four
input states with average fidelities of $0.695 \pm 0.001$, $0.676 \pm 0.005$, $0.697 \pm 0.001$ and
$0.681 \pm 0.007$, respectively. All of the experimental
fidelities are significantly higher than that of a classical measure-and-prepare (M\&P) cloner ($F_\mathrm{M\&P}=0.5$), where the clones are prepared from a dual-homodyne measurement of an input state \cite{grosshans2001quantum}. More importantly, all the clones also surpass the no-cloning limit of $F_\textrm{2}=2/3$, which is impossible even with a perfect deterministic cloning machine. 

To further analyse the HCM, we examine the cloning of
an input state $(x,p)=\left(2.23, 2.19\right)$ ($|\alpha|= 1.56$) in greater detail. This experiment is repeated \num{5e7} times, from which about \num{5.9e5} runs produced successfully heralded clones. The electronic gain $g_\textrm{x,p}$ and the splitting ratio of the beam
splitter are carefully tuned to ensure that the two clones produced
are nearly identical. 
The probabilistic heralding function was
chosen to ensure that the output clones have exactly unity gain on average. This is done to
prevent any overestimation of the fidelity (see Supplementary Note 3 and Supplementary Fig. 1). As can be
seen in Fig.~\ref{fig:2}c, the produced clones have noise
significantly lower than the M\&P cloning protocol. Since the noise variance
is only affected by the deterministic amplification,
setting $g_\textrm{NLA}>1$ will reduce the required DLA gain while still
achieving unity gain. As a result, the clones produced by our HCM will
have less noise compared to its deterministic counterpart. The data points also show that the heralded events (purple region of Fig.~\ref{fig:2}c right) have Gaussian distributions with mean equal to that of the input state.  We experimentally obtained a fidelity of $0.698 \pm 0.002$ and $0.697 \pm 0.002$ for the two clones. The fidelity plot in
Fig.~\ref{fig:2}d clearly demonstrates that the fidelities of both
clones exceed not only the M\&P limit but also lie beyond the
no-cloning limit by more than 15 standard deviations.\\

\begin{figure*}[t]
\begin{center}
\includegraphics[width = 16cm]{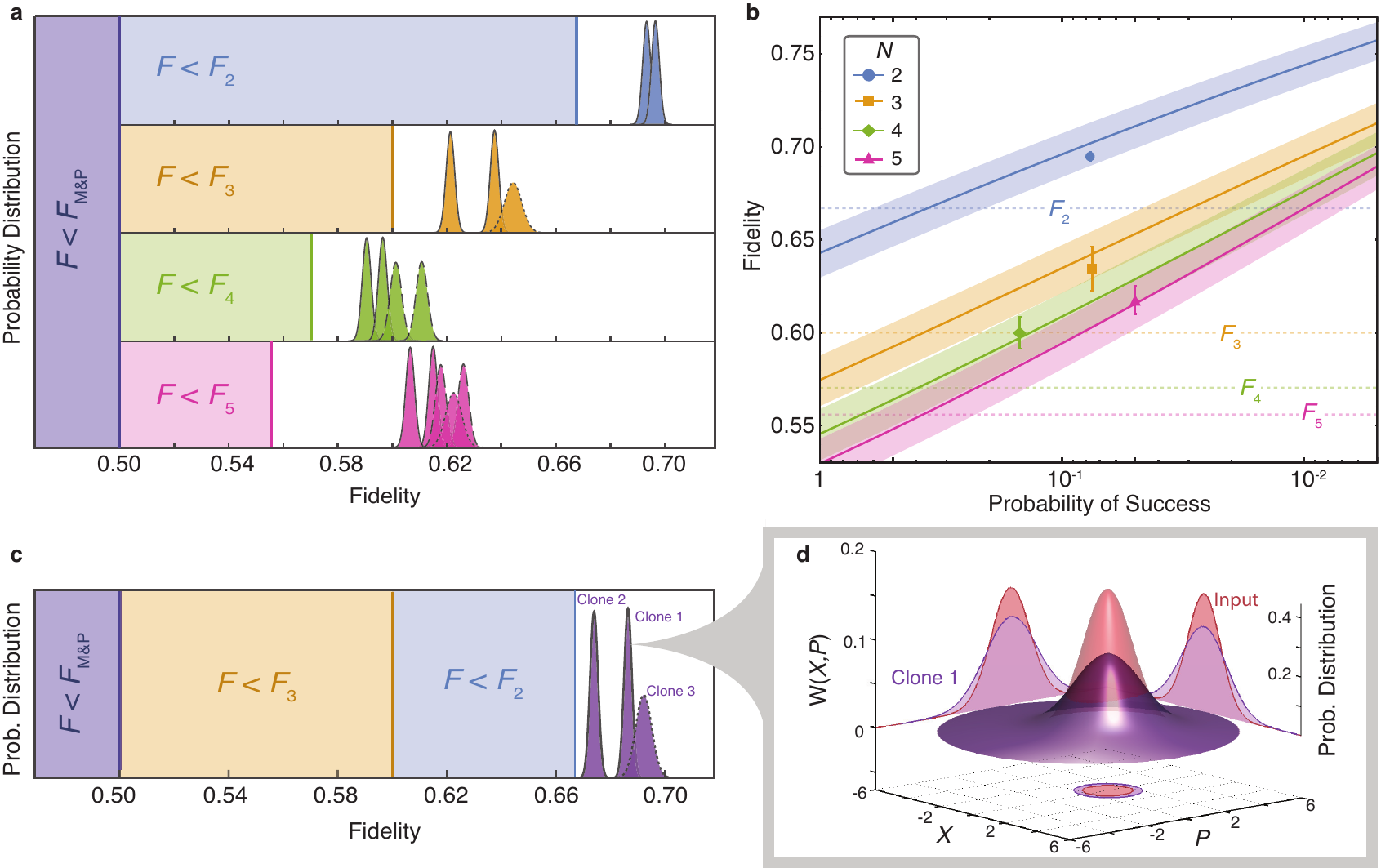}
\caption{\textbf{Multiple clones with hybrid cloning machine.} ({\bf a}) Fidelity of $N$
  clones beyond the no-cloning limit. By applying appropriate
  deterministic and probabilistic gains on the input
  $|\alpha|\approx 0.5$, clones with fidelity
  exceeding their corresponding no-cloning limits $F_N=N/(2N-1)$ are produced. For $N>2$, only two of the output clones are
  directly measured (solid lines). The remaining $N-2$ clones'
  fidelity distributions are obtained either from
  rescaled data of different runs (dashed) or estimation of the
  remaining intensities (dotted). A sample
  size of $5\times 10^7$ data points is used for all $N$. The spreads in fidelity distributions are predominately
due to imperfect splitting. ({\bf b}) Fidelity as a
  function of heralding probability of success for different $N$. Theoretical simulations (solid lines) are superimposed with the experimental points (symbols) and the no-cloning limits $F_N$ (dotted lines). Error bars represent 1 standard deviation of clones' fidelities and the shaded regions are theoretical expected fidelities from 1 standard deviation of the dual-homodyne detection efficiency. ({\bf c}) Three clones
  with fidelity $F>F_2$ and $F_3$. ({\bf d}) The experimentally reconstructed Wigner functions
  of the input (red) and clone 1 (purple) together with their normalized
  probability distributions for both $X$ and $P$ quadratures. The quadrature values are normalized to the vacuum. \label{fig:3} }
\end{center}
\end{figure*}

\noindent \textbf{Multiple Clones.} We operate our HCM at higher gains to enable the production of more than two clones. \blue{In order to have $g_\text{NLA}>1$, from equations~\eqref{eq:ug} and~\eqref{eq:ts}, we require $g_\text{DLA}<\sqrt{N}$, which leads to $T_\text{s}>1/N$. Hence, by tailoring $T_\text{s}$ for each $N$, HCM can produce $N$ clones with
fidelity beating the deterministic bound
$F_{N}$ with the desired probability of success.} Fig.~\ref{fig:3}a shows the fidelity of the
multiple clones with an input of $|\alpha|\approx 0.5$. The average fidelities of
the clones for $N=2, 3, 4$ and $5$ are $0.695 \pm 0.002$,
$0.634 \pm 0.012$, $0.600 \pm 0.009$ and $0.618 \pm 0.007$,
respectively, clearly surpassing the corresponding no-cloning limit. In Fig.~\ref{fig:3}b, we plot the theoretical prediction of the fidelity as a function of the probability of success with the experimental data. The theoretical fidelity is modelled upon the dual-homodyne detection efficiency of $90 \pm 5\%$, which is the main source of imperfection (see Supplementary Note 3 for details). We find that our results lie well within the expected fidelities, with the probability of success ranging between $5\%$ to $15\%$. Remarkably, by keeping $5\%$ of the data points, the average cloning fidelity for $N=5$ can be enhanced by more than $15\%$, and hence exceeding the no-cloning limit $F_5$ by $11.2\%$. 

For deterministic unity gain cloners, as long as $N$ clones are produced each with fidelity $F>F_{N+1}$~\cite{Braunstein2001,fiuravsek2001optical}, 
one may conclude that there are no other clones with equal or
higher fidelity. Here we show that this is not necessarily the case for probabilistic cloning. By further increasing NLA gain, we successfully
produce three clones, each with fidelity $F>F_2$ (Fig.~\ref{fig:3}c), and the average fidelity is $0.684 \pm 0.009$. Given only fidelity, it is impossible for a receiver with only two clones to determine whether the clones originate from a 2-clone or 3-clone probabilistic protocol (Fig.~\ref{fig:3}a and \ref{fig:3}c). The resulting probability distribution from 
  \num{7.2e6} successful three-clone states out of
  \num{5e8} inputs and the corresponding experimental reconstructed Wigner function are shown in Fig.~\ref{fig:3}d together with the input state. 

The theoretical fidelity for the HCM's clones at unity gain can be shown to be $F_\text{HCM}=1/(1+(g_\textrm{DLA}^2-1)/N)$, which is only a function of the deterministic gain and the number of clones. We note that maximum fidelity for a given $N$ can be achieved in the limit of $T_\text{s}\rightarrow 1$, giving $F_\text{max}(N)=1/(1+(\sqrt{N}-1)/N)$ (See Supplementary Note 3). $F_\text{max}(N)$ converges to 1 in the limit of an infinite number of clones. However, since this also requires an infinitely large nondeterministic gain, and thus an unbounded truncation in post-selection, the probability of success will be essentially zero.

\section{Discussion}
In summary, we have proposed and demonstrated a hybrid cloning machine that combines a deterministic 
and a probabilistic amplifier to clone unknown coherent states with fidelity beyond the no-cloning limit.
Even though an ideal NLA implementation is not possible with our setup, as this would require zero deterministic gain, our hybrid approach does allow the integration of measurement-based NLA in the optimal deterministic amplifier. We showed that our device is capable of high-fidelity cloning of large coherent states and generation of
multiple clones beyond the no-cloning limit, limited only by the amount of data collected and
the desired probability of success. Our cloner, while only working probabilistically, provides
a clear heralding signal for all successful cloning events.

Several comments on the prospects and avenues for future work are in order. An immediate extension is the implementation of HCM in various feed-forward based cloning protocols, such as phase conjugate cloning~\cite{sabuncu2007experimental}, cloning of Gaussian states~\cite{olivares2006cloning,weedbrook2008quantum}, telecloning~\cite{koike2006demonstration}, and cloning with prior information~\cite{Muller2012,sacchi2007phase,sabuncu2008experimental}. Our tunable probabilistic cloner could further elucidate fundamental concepts of quantum mechanics and quantum measurement, for instance, quantum deleting~\cite{qiu2002combinations} and quantum state identification~\cite{chefles1999strategies,Shehu2015}. This probabilistic coherent protocol might also play a role in the security analysis of eavesdropping attacks in continuous variable quantum cryptography as well~\cite{grosshans2004continuous,weedbrook2006coherent}. The implication of HCM in the context of quantum information distributor~\cite{braunstein2001quantum} and quantum computation~\cite{galvao2000cloning} also demands further investigation.

Beyond probabilistic cloning, owing to the robustness and ease of implementation of this heralded hybrid amplification, we envisage numerous
applications in quantum communication~\cite{Chrzanowski2014,Ulanov2015,combes2016models}, quantum teleportation~\cite{fuwa2014noiseless,He2015,RBlandino2016} and quantum error correction~\cite{Ralph2011error}. As such, we believe our scheme will be a useful tool in the quest to realise large-scale quantum networks.

\section{Methods} 
\noindent \textbf{Experimental Details.} Our hybrid cloning machine is shown in
Fig.~\ref{fig:1}b. The coherent state is created by modulating the
sidebands of a \SI{1064}{nm} laser at \SI{4}{MHz} with a pair of
phase and amplitude modulators. In the cloning stage, the input
mode is split by a variable beam splitter consisting of a half-wave
plate and polarising beam splitter with transmissivity
$T_\text{s}=(g_\textrm{NLA}/g_\textrm{DLA})^2$. An optical dual-homodyne
measurement is performed on the reflected beam, where the measurement
outcome is further split into two parts. The first part is used to
extract the \SI{4}{MHz} modulation by mixing it with an electronic
local oscillator, before being low pass filtered at \SI{100}{kHz} and
oversampled on a 12-bit analog-to-digital converter at \SI{625}{kSamples} per second. The data is
used to provide the heralding signal. The second part of the output is
amplified electronically with a gain $g_\textrm{x,p}$ and sent to another
pair of phase and amplitude modulators, modulating a bright auxiliary
beam. This beam is used to provide the displacement operation by
interfering it in phase with the delayed transmitted beam on a 98:2
beam splitter. The delay on the transmitted beam ensures that it is
synchronised to the auxiliary beam at the beam splitter. The combined
beam is then split by an $N$-port splitter to generate clones. The
clones are then verified individually by the same homodyne
detector. Two conjugate quadratures $X$ and $P$ are recorded and used
to characterise the Gaussian output. For each separate homodyne
detection at least $5\times 10^7$ data points are saved. We note that in evaluating the fidelities, we take into account the detection efficiency and losses to avoid an overestimation of the fidelity (see Supplementary Note 3).\\

\noindent\textbf{Acknowledgements}
\newline
We thank S. Armstrong for helpful discussions. The research is supported by the Australian Research Council (ARC) under the Centre of Excellence for Quantum Computation and Communication Technology (CE110001027). PKL is an ARC Laureate Fellow.
\newline
\newline
\textbf{Author contributions}
\newline
JD, RB, SMA and TCR developed the theory. JYH, JZ, SMA, MB, TS and PKL conceived and conducted the experiment. JYH, JZ, SMA analysed the data. JYH, JZ, MB and SMA drafted the initial manuscript. TS, TCR and PKL supervised the project. All authors discussed the results and commented on the manuscript. The authors declare no competing financial interests.\\

\setcounter{section}{0}
\renewcommand{\figurename}{Supplementary Figure}
\setcounter{figure}{0}
\setcounter{equation}{0}
\renewcommand{\thefigure}{S\arabic{figure}}
\renewcommand{\theequation}{S\arabic{equation}}
\begin{center}
\large{\textbf{Supplementary Material}}
\end{center}
\vskip-35cm
\section*{Supplementary Note 1: Equivalent Theory for Hybrid Cloning Machine}
\label{supp:eq}
Here we show how concatenating a noiseless linear amplifier (NLA) and deterministic linear amplifier (DLA) (Fig. 2a) can form the amplification stage of an amplify-and-split hybrid cloning machine (HCM). In our scheme, the deterministic amplification can be implemented by a linear optical
feed-forwarding circuit~\cite{Andersen2005}, which is shown in Fig. 2b. A beam splitter with
transmission
\begin{equation}
  \label{eq:1}
T=\frac{1}{g_{\mathrm{DLA}}^2},\;
\end{equation}
is used to tap off the noiselessly amplified state. The reflected beam is subjected to
a dual homodyne measurement whose outcome $d=(x,p)$ is electronically amplified
with gain
\begin{equation}
  \label{eq:2}
  g_{\mathrm{e}}^{\mathrm{(ff)}} = \sqrt{2\left( g_{\mathrm{DLA}}^2-1 \right)}.
\end{equation}
This amplified signal is feed-forwarded to the transmitted beam to displace it by $g_{\mathrm{e}}^{\mathrm{(ff)}}d$ via electro-optical modulators (EOMs). 

As described in~\cite{Remi2015}, this same
output state can be obtained by a different setup (Fig. 2c) where the NLA is moved from the input to the reflected port with a modified gain
\begin{equation}
  \label{eq:3}
  g_{\mathrm{NLA}'} = \sqrt{\frac{1-T}{1-T g_{\mathrm{NLA}}^2}}g_{\mathrm{NLA}},
\end{equation}
and the beam splitter is replaced by a beam splitter
with transmission
\begin{equation}
  \label{eq:4}
  T_\text{s} = T g_{\mathrm{NLA}}^2=g_{\mathrm{NLA}}^2/g_{\mathrm{DLA}}^2.
\end{equation}
These setup are equivalent provided $g_{\mathrm{NLA}} < g_{\mathrm{DLA}}$. The reason for going to this alternate
setup is that now we have a situation where the NLA is followed by a dual-homodyne detection which can be implemented by a measurement-based NLA (MBNLA)~\cite{Fiurasek2012,Chrzanowski2014}. The MBNLA consists of a Gaussian heralding function $p_\text{F}$, followed by a rescaling factor
\begin{equation}
  \label{eq:6}
  g_{\mathrm{e}}^{\mathrm{(rescale)}}=\frac{1}{g_{\mathrm{NLA}'}}.\;
\end{equation}
This rescaling factor is combined with $g_{\mathrm{e}}^{\mathrm{(ff)}}$ to give a net electronic gain of
\begin{align}
  \label{eq:7}
  g_{\mathrm{x,p}}  &=g_{\mathrm{e}}^{\mathrm{(rescale)}}  g_{\mathrm{e}}^{\mathrm{(ff)}}\nonumber\\
  &=\sqrt{2\left(\frac{1}{T_\text{s}}-1 \right)},
\end{align}
as shown in Fig. 2e. Finally, subjecting the output to an $N$-port beam splitter results in a total gain
\begin{equation}
\label{eq:unitgain}
g=g_{\mathrm{NLA}}g_{\mathrm{DLA}}/\sqrt{N}
\end{equation}
Cloning is achieved when the gain is set to unity, i.e.~$g=1$. Upon setting $T_\text{s}$ and the number of clones $N$, the corresponding $g_{\mathrm{NLA}}$ and $g_{\mathrm{DLA}}$ at unity gain can be determined from Supplementary equation~\eqref{eq:4}. Combining Supplementary equations~\eqref{eq:4} and \eqref{eq:unitgain}, we note that as long as $T_\text{s}>1/N$, $g_{\mathrm{NLA}}$ will always be bigger than 1, enabling the hybrid operation of the cloning machine.

\section*{Supplementary Note 2: Implementation of Measurement-Based Noiseless Linear Amplifier}
\label{supp:mbnla}
Here we give a brief summary of the MBNLA implementation in our protocol. The NLA with
gain $g_{\text{NLA}'}$ on the reflected mode followed by a dual-homodyne
measurement can be replaced by a direct dual-homodyne measurement on the
reflected mode, whose outcomes $\alpha_\text{m}=(x_\text{m}+ip_m)/\sqrt{2}$ are used to herald successfully cloned states. These measurement data points $\alpha_\text{m}$ are accepted with probability
\begin{align}
\label{eq:filter}
  p_\text{F}\left(\alpha_\text{m} \right) =
\begin{cases}
  \frac{1}{M}\exp\left[|\alpha_\text{m}|^2\left(1-\frac{1}{g_{\text{NLA}'}^2} \right) \right]  &\mbox{if } |\alpha_\text{m}| < |\alpha_\text{c}|\\
  1 &\mbox{otherwise.}
\end{cases}
\end{align}
Here  $|\alpha_\text{c}|$ is the NLA cut-off and $M=\exp \left[|\alpha_\text{c}|^2 \left(1-1/g_{\text{NLA}'}^2 \right) \right]$ ensures that the output is normalized properly. This heralding function, $p_\text{F}\left(\alpha_\text{m} \right)$, together with the rescaling factor (Supplementary equation~\eqref{eq:6}), can be made arbitrarily close to an ideal NLA operation $g^{n}$, where $n$ is the number operator.

Next, we describe how the parameters $g_{\text{NLA}'}$ and $|\alpha_\text{c}|$ are chosen in Supplementary equation~\eqref{eq:filter}. For both $X$ and $P$ quadratures, the gain $g_{\text{NLA}'}$ is chosen such that the inferred mean of $N$ clones prior to splitting equals to that of the amplified input mean to ensure average unity gain. For the quadrature with zero mean, $g_{\text{NLA}'}$ will be tuned such that the variance matches the value given by the experimental model with imperfect dual-homodyne detection efficiency. In all cloning protocols, only two clones are directly measured. Based on the $N$-port splitting ratios, the mean and the variance of the
remaining clones for $N>2$ can be evaluated either from rescaled data from different cloning runs or from an estimation of the remaining transmission power.

\jy{The cut-off parameter $|\alpha_\text{c}| > 0$ determines how closely the MBNLA
approximates an ideal NLA. A larger cut-off parameter implements the
ideal NLA more accurately at the cost of a lower probability of
success.  The accuracy and the success probability also depend
on the $g_{\text{NLA}'}$ and the amplitude of the coherent
state. The probability distribution of a dual-homodyne measurement on $\rho=\ket{\alpha_0}\bra{\alpha_0}$ is given by
\begin{eqnarray}
Q(\alpha_\text{m}) &=& \frac{1}{\pi}\left<\alpha_\text{m} |\rho| \alpha_\text{m} \right>\nonumber \\
&=& \frac{1}{\pi} \exp(-|\alpha_\text{m}-\alpha_0|^2),
\end{eqnarray}
which is centred around $\alpha_0$ with the variances $\text{Var}(\text{Re}(\alpha_\text{m}))=\text{Var}(\text{Im}(\alpha_\text{m}))=0.5$. Applying the probabilistic filter Supplementary equation~\eqref{eq:filter} on the distribution $Q(\alpha_\text{m})$ results in a two-dimensional Gaussian distribution with amplified mean and variance of $g_{\text{NLA}'}^2 \alpha_0$ and $0.5g_{\text{NLA}'}^2$, respectively. To implement the NLA with high fidelity, we propose the following cut-off value for the distribution:
\begin{equation}
 \text{Re}(\alpha_\text{c})={g^2_{\text{NLA}'}}\text{Re}(\alpha^\text{max}_0)+\beta(\sqrt{0.5}g_{\text{NLA}'})
\end{equation}
and similarly for $\text{Im}(\alpha_\text{c})$. Here, $\alpha^\text{max}_0$ is the expected maximum input amplitude involved in the cloning protocols. In our experiment, $\beta$ is chosen to ensure more than $98\%$ of the data are within the cut-off value for both the two-clone and multi-clone protocols.}

Finally, the probability of success for input state $\ket{\alpha_0}$ can be obtained by integrating the function $p_\text{F}\left(\alpha_\text{m} \right)$ (Supplementary equation~\eqref{eq:filter}) with the dual-homodyne distribution $Q(\alpha_\text{m})$.

\section*{Supplementary Note 3: Fidelity Evaluation}
\label{supp:fid}
\begin{figure}[t]
\centering
\includegraphics[width=8cm]{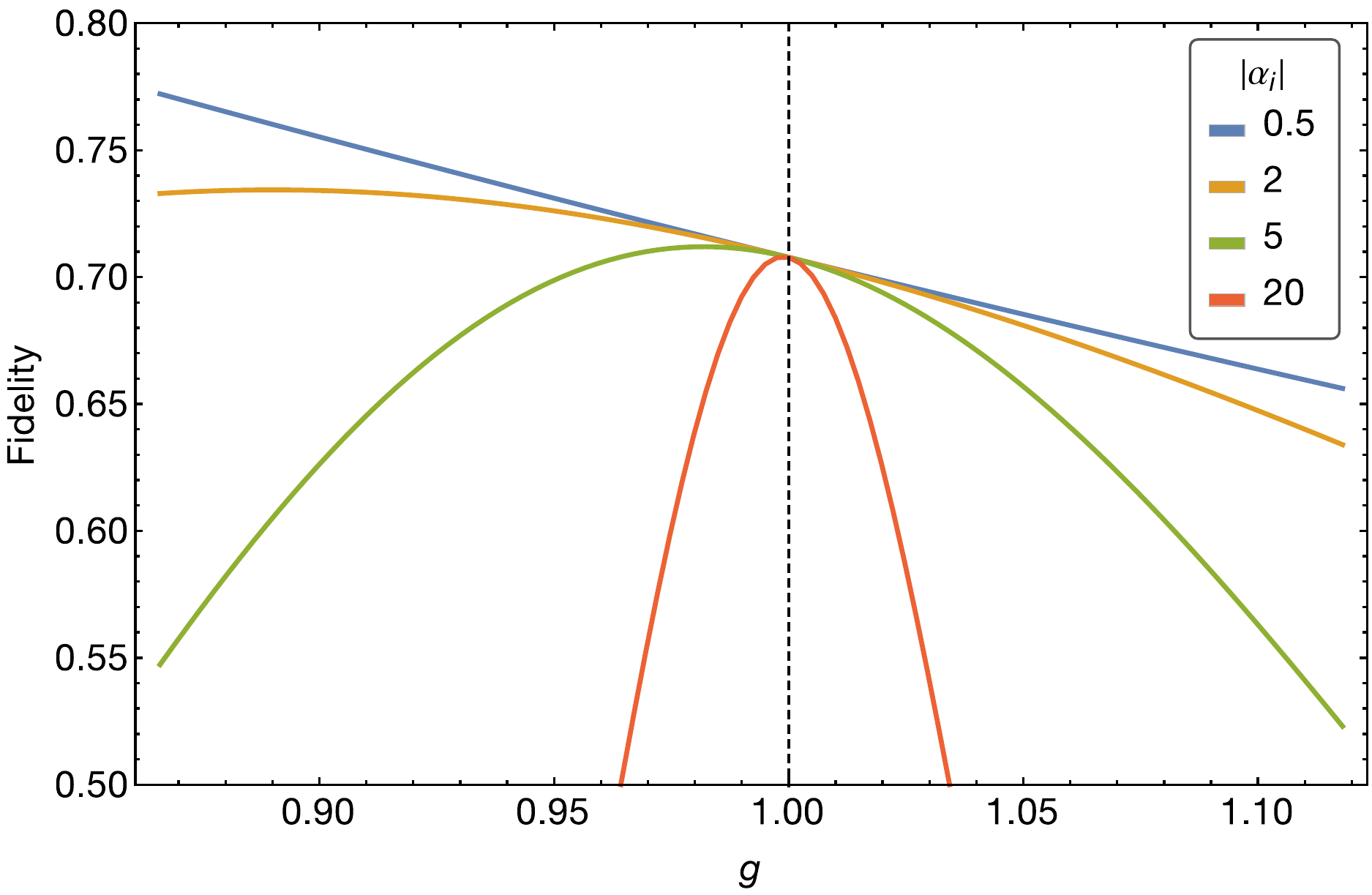}
\caption{\label{fig:fid} \textbf{Fidelity of two-clone protocol with different input states.} Deviation from unity gain (dashed line) can lead to overestimation of fidelity when the amplitude of the input state $|\alpha_\text{i}|$ is small.}
\end{figure} 
The fidelity which quantifies the overlap between the clones and the
input state is calculated as a criterion for examining the performance
of our HCM. We consider single-mode Gaussian input
$\rho_{\text{i}}\left( \textbf{d}_{\text{i}},
  \textbf{V}_{\text{i}}\right)$
and output
$\rho_{\text{o}}\left( \textbf{d}_{\text{o}},
  \textbf{V}_{\text{o}}\right)$,
where $\textbf{d}_j=(x_j,p_j)$ is the mean of the amplitude and the
phase of the state $\rho_j$ while
$\textbf{V}_j=\text{diag}(\sigma_{x_j},\sigma_{p_j})$ is the corresponding covariance matrix. The fidelity between $\rho_{\text{i}}$ and
$\rho_{\text{o}}$ is given by~\cite{weedbrook2012gaussian}
\begin{align}
F(\rho_{\text{i}},\rho_{\text{o}})=\frac{2}{\sqrt{\vartriangle+\delta}-\sqrt{\delta}}\text{exp}\left[ -\frac{1}{2}{\bf d}^{\it{T}}({\bf V}_{\text{i}}+{\bf V}_{\text{o}}) ^{-1} {\bf d}\right]\;,
\end{align}
where $\vartriangle :=\text{det}(\bf{V}_{\text{i}}+\bf{V}_{o})$, $\delta :=(\text{det} {\bf{V}_{\text{i}}}-1)(\text{det} {\bf{V}_{\text{o}}}-1)$, and $\bf{d}$ $:=\textbf{d}_{\text{o}}- \textbf{d}_{\text{i}}$. The fidelity for a coherent state input $\rho_{\text{i}}\left( \textbf{d}_{\text{i}}, \textbf{1}\right)$ is
\begin{eqnarray}
F&=&\frac{2}{\sqrt{(\sigma_{x_\text{o}}^2+1)(\sigma_{p_\text{o}}^2+1)}}\nonumber\\
& &\text{exp}\left[ -\frac{1}{2} \left(\frac{(x_{\text{o}}-x_{\text{i}})^2}{\sigma_{x_\text{o}}^2+1} + \frac{(p_{\text{o}}-p_{\text{i}})^2}{\sigma_{p_\text{o}}^2+1}\right)\right] \;.
\label{eq:fid}
\end{eqnarray}
Suppose that the output quadratures are symmetric, and the theoretical output means $\{x_\text{o},p_\text{o}\}$ are given by $\{g x_\text{i},g p_\text{i}\}$ and the variances $\sigma_{x_\text{o}}^2=\sigma_{p_\text{o}}^2=1+2(g_\text{DLA}^2-1)/N$, respectively. Then, the theoretical fidelity of $N$ clones is thus
 \begin{equation}
 F(N)=\frac{1}{1+(g_\text{DLA}^2-1)/N}\nonumber\text{exp}\left[ -\frac{(g-1)^2|\alpha_\text{i}|^2}{1+(g_\text{DLA}^2-1)/N}\right] \;.
\label{eq:fidN}
 \end{equation}
 where $\alpha_\text{i}=(x_\text{i}+ip_\text{i})/2$. We emphasize that it is crucial to set the gain of the HCM as close to unity gain as possible. As shown in Supplementary Fig.~\ref{fig:fid}, non-unity gain for small input amplitude may lead to an overestimation of the fidelity. At unity gain, the theoretical fidelity reduces to
\begin{equation}
F(N)=\frac{1}{1+(g_\text{DLA}^2-1)/N}\;,
\end{equation}
which depends only on the deterministic gain and number of clones. The maximum fidelity can be achieved in the limit of $T_\text{s}\rightarrow 1$, and from Supplementary equation~\eqref{eq:4} and \eqref{eq:unitgain}, $g_\text{DLA}^2\rightarrow \sqrt{N}$, giving
\begin{equation}
F(N)\rightarrow F_\text{max}(N)=\frac{1}{1+(\sqrt{N}-1)/N}\;.
\end{equation}
\jy{Note that although the fidelity goes to 1 as $N$ increases, the probability of success is vanishingly small. This is because at this limit, the cut-off $\alpha_\text{c}$ in the heralding function will also scale with the probabilistic gain $g_\text{NLA}^2$ to implement the amplification faithfully, thus rejecting essentially most of the data points (c.f.\ Supplementary equation~\eqref{eq:filter}).} 

In practice, the fidelity of HCM is limited by several factors, such as the dual-homodyne efficiency, electronic gain, asymmetry in the quadratures and imperfect $N$-port splitting ratio. Nevertheless, following~\cite{andersen2007optical}, a simple model of the imperfection of HCM can be constructed simply by considering the detection efficiencies of the dual-homodyne $\eta_{\text{DH}}$, which is the dominant source of imperfection. Taking into account the losses and imperfect visibilities, the average dual-homodyne detection efficiency for the amplitude and phase quadratures is $90\pm5\%$.

To evaluate the experimental fidelity of the clones, we corrected the homodyne data to ensure proper characterization of both the input state and the clones. 
This is to avoid overestimation of the fidelity due to underestimation of the clones variances (See Supplementary Fig.~\ref{fig:fid}). The  total detection efficiency for the input state is typically around $97\%$, where we have taken into account the quantum efficiency of the photodiodes, mode-matching visibility, and the propagation losses. The detection efficiency of the clones in                                                                                                                                                                                                                                                                                                                                                                                                                                                                                                                                                                                                                          the verification stage is about $98.5\%$. 

The correct mean can be obtained by rescaling the overall data by $1/\sqrt{\eta_\textrm{tot}}$, where
$\eta_\textrm{tot}$ is the total detection efficiency of either the input state or the clones. The corresponding variance can be obtained by subtracting $(1-\eta_\textrm{tot})/\eta_\textrm{tot}$ from the variance of the rescaled data.

To determine the standard deviation of the fidelity, we take into account the propagation of the uncertainties in the variance of output quadratures. The uncertainty of our fidelity is estimated according to: 
\begin{align}
\textrm{Var}\left( \sigma_{\tilde{F}} ^2\right) = \left( \frac{\partial \tilde{F}}{\partial \sigma_{x_\text{o}}^2} \right)^2 \textrm{Var}(\sigma_{x_\text{o}}^2) + \left( \frac{\partial \tilde{F}}{\partial \sigma_{p_\text{o}}^2} \right)^2 \textrm{Var}(\sigma_{p_\text{o}}^2) \;.
\end{align} 
Here $\tilde{F}$ is 
\begin{align}
\tilde{F}=\frac{2}{\sqrt{(\sigma_{x_\text{o}}^2+1)(\sigma_{p_\text{o}}^2+1)}},
\label{eq:fidu}
\end{align} 
which is obtained by setting $x_{\text{o}}= x_{\text{i}}$ and $p_{\text{o}}=p_{\text{i}}$ in Supplementary equation~\eqref{eq:fid}. The variance of the quadrature variances $\textrm{Var}(\sigma_{x_\text{o}}^2)$ and $\textrm{Var}(\sigma_{p_\text{o}}^2)$ are evaluated based on several parameters: number of datapoints, uncertainty of the total detection efficiency and uncertainty of the $N$-port beam splitter. Finally the standard deviations of the average fidelities are determined from the spread of the fidelity probability distributions.

\bibliographystyle{naturemag}

\end{document}